\documentclass[preprint,aps,floatfix]{revtex4-1}
\usepackage{hyperref}
\everymath={\displaystyle}
\usepackage{young}
\usepackage{rotating}
\usepackage{longtable}
\def\1{\mbox{l\hspace{-0.53em}1}}

\usepackage{multirow}

\newlength{\AccoHaut}

\begin{document}
\title{$[{\bf 70, \ell^+}]$ baryons in large $N_c$ QCD revisited:
The effect on Regge trajectories.}

%%%\author{N. Matagne\address{University of Li\`ege, Institute of Physics B5, Sart Tilman,
%%%B-4000 Li\`ege 1, Belgium}\thanks{e-mail address: nmatagne@ulg.ac.be} and Fl. Stancu\addressmark\thanks{e-mail address: fstancu@ulg.ac.be}} 

\author{N. Matagne\footnote{E-mail address: nicolas.matagne@umons.ac.be}}
\affiliation{University of Mons, Service de Physique Nucl\'eaire et 
Subnucl\'eaire,Place du Parc 20, B-7000 Mons, Belgium}

\author{Fl. Stancu\footnote{E-mail address: fstancu@ulg.ac.be}}
\affiliation{University of Li\`ege, Institute of Physics B5, Sart Tilman,
B-4000 Li\`ege 1, Belgium}

\date{\today}

%\begin{document}
%\maketitle

\begin{abstract}
A simple procedure within the $1/N_c$ expansion method where
all the $N_c$
quarks are treated on the same footing has been found successful in  
describing mixed symmetric negative parity baryon states belonging to the $[{\bf 70},\ell^-]$ multiplets
%with $\ell$ = 1, 2 or 3. 
of the  $N$ = 1 and 3 bands.
Presently it is applied
to mixed symmetric positive parity $[{\bf 70},0^+]$ and $[{\bf 70},2^+]$ multiplets
of the $N$ = 2 band.
We search for the most dominant terms in the mass formula. The results are compared 
to those obtained in the procedure where the system is separated into  
a  core and an excited quark. 
We find that both the spin and isospin operators of the entire system of $N_c$ quarks 
play  dominant roles in describing the data, like for negative parity states.
As a by-product we present the contribution
of the leading spin-isospin singlet term 
as a function of the band number, which hints at distinct Regge 
trajectories for the symmetric and mixed symmetric states.
\end{abstract}

\maketitle

\section{Introduction}

The $1/N_c$ expansion method, where $N_c$ is the number of colors  \cite{HOOFT,WITTEN}, 
is based on the discovery
that, for $N_f$ flavors, the ground state baryons
display an exact SU($2N_f$) spin-flavor symmetry in the
large $N_c$ limit of QCD \cite{DM93}. 
Presently it is considered to be a model independent, powerful and  systematic tool for
baryon spectroscopy. 
It has been applied with great success
to the ground state baryons ($N = 0$ band), described
by the symmetric representation $\bf 56$ of SU(6), where $N_f$ = 3
\cite{DM93,Jenk1,DJM94,DJM95,CGO94,JL95,DDJM96}. 
At $N_c \rightarrow \infty$ the ground state baryons are degenerate.
At large, but finite $N_c$, the mass splitting 
starts at order $1/N_c$. 

The extension of the $1/N_c$ expansion method
to excited states is based on the observation these states can approximately be classified  
as  SU($2N_f$) multiplets, and that the resonances can be grouped into 
excitation bands, $N $ = 1, 2, ..., as in quark models,
each band containing a number of SU(6) $\times$ O(3) multiplets. 
The symmetric multiplets of these bands were analyzed by analogy 
to the ground state. In this case the splitting starts at order $1/N_c$ as well.

The study of mixed symmetric multiplets was less straightforward,
being technically more complicated.  
Two procedures have been proposed and applied to the     
excited states belonging to the $[{\bf 70},1^-]$ multiplet ($N = 1$ band).
The first one is based on the separation of the system into a ground state core + an 
excited quark
\cite{CGKM,Goi97,PY,CCGL,CaCa98,SGS,Pirjol:2003ye}. It is an extension of the 
ground state treatment to excited states inspired by the Hartree picture. 
Later on it was supported by the authors of Ref.  \cite{Pirjol:2007ed}.
In the second method, proposed by us  \cite{Matagne:2006dj}, the system of $N_c$  
quarks is treated as a whole. All identical quarks are considered 
on the same footing and therefore the Pauli principle is satisfied.
It has successfully been applied to the negative parity multiplet $[{\bf 70},1^-]$ of
the $N$ = 1 band \cite{Matagne:2011fr}
and recently  to the multiplets $[{\bf 70},\ell^-]$ (with $\ell$ = 1,2,3) of 
the $N = 3$ band 
\cite{Matagne:2012tm}. The advantage is that our mass formula has fewer terms than the 
ground state core + excited quark method, so that it is physically more transparent.

It is worth mentioning that in both procedures the mass splitting of mixed symmetric states starts 
at order $N^0_c$ and they are both compatible with the meson-nucleon scattering picture
\cite{HAYASHI,MAPE,MATTIS,MattisMukerjee}. For the ground state core + excited quark approach this 
has been shown in Refs. \cite{Pirjol:2003ye,COLEB1} by using a mass formula with three leading 
operators (one of order $N_c$, two of order $N^0_c$) generating three sets of degenerate 
states called {\it three~towers~of~states} \cite{Pirjol:2003ye} for $\ell$ = 1. In Ref. \cite{Matagne:2011sn}
we gave an explicit proof of the degeneracy of mass eigenvalues for $\ell$ = 3.
In a similar way, but with three
different leading operators in our approach, we have also proven the above compatibility for mixed symmetric 
$\ell$ = 1  \cite{Matagne:2012vq} states in SU(4). 
Note that in both procedures only operators containing components of the angular momentum start
at order $N^c_0$.

Here we wish to test our method \cite{Matagne:2006dj} on mixed symmetric positive
parity multiplets, studied so far within the ground state core + excited quark approach
\cite{Matagne:2005gd,Matagne:2006zf}. Our study is especially motivated by the fact that a recent
 multichannel partial wave 
analysis has revealed the existence of new positive parity resonances which appeared in
the 2012 version of the Review of Particle Properties (PDG) \cite{PDG}. 

The paper is organized as follows. In the next section we shortly review  previous studies on positive parity
resonances within the symmetric core + excited quark method and present the orbital-flavor-spin wave
function in the new approach
\cite{Matagne:2006dj}. In Section III we introduce the mass 
operator and derive the analytic expressions of the matrix elements of the required operators as a 
function of $N_c$, using the method of 
Ref. \cite{Matagne:2006dj}. In Section IV the experimental situation is shortly reviewed.
In Section V our results for mixed symmetric positive parity states are presented. 
Section VI is devoted to Regge trajectories and the last section contains some conclusions.
%%%%%%%%%%%%%%%%%%%%%%%%%%%%%%%%%%%%%%%%%%%%%%%%%%%%%%%%%%%%%%%%%%
\section{The lowest excited positive parity states}

The lowest excited positive parity resonances belong to the 
$N = 2$ band, which contains the multiplets $[{\bf 56'},0^+], [{\bf 56},2^+],
[{\bf 70},0^+], [{\bf 70},2^+]$ and $[{\bf 20},1^+]$. 
Radially excited states $[{\bf 56'},0^+]$ have been studied in Ref.  \cite{CC00}
with a simplified - G\"ursey-Radicati type - mass formula.
The masses of the baryons
supposed to belong to  the multiplet 
$[{\bf 56},2^+]$ 
%or $[{\bf 56'},0^+]$ 
have been calculated within the $1/N_c$ expansion method
in Ref. \cite{GSS03}.  
The approach of Ref. \cite{GSS03} has been extended to
higher excitations  belonging to the $[{\bf 56},4^+]$ multiplet
($N=4$ band) \cite{MS1}. 
We recall that in the symmetric representation it is not
necessary to distinguish between excited and core quarks, 
thus the wave function has a simple structure \cite{GSS03}.

As already mentioned, the mixed symmetric multiplets $[{\bf 70},0^+]$ and $[{\bf 70},2^+]$ have been studied 
by applying the symmetric core + an excited quark approach  \cite{Matagne:2005gd,Matagne:2006zf}. 
The structure of the 
intrinsic orbital wave function was rather complicated,
containing a term with an excited symmetric core in contrast to the 
wave function of the $N$ = 1 band, where the symmetric core was in the ground state \cite{CCGL}.
Such a wave function has been constructed in Ref. \cite{Matagne:2005gd} by using   
generalized Jacobi coordinates  \cite{MOSHINSKY}  and  fractional parentage techniques \cite{Stancu:1991rc}.
It had been first applied to SU(4) ($N_f$ = 2) \cite{Matagne:2005gd}
and next extended to SU(6) ($N_f$ = 3) baryons \cite{Matagne:2006zf}.

Presently we treat the system of $N_c$ quarks as a whole, no quark separation.
Both the orbitally excited and the spin-flavor parts of the total wave function are described by the partition 
$[f]$ = $[N_c - 1, 1]$. By inner product rules of the permutation group one can form a totally symmetric
orbital-spin-flavor wave function  described by the partition  $[N_c]$. 
Following Ref. \cite{Matagne:2011fr} the most general form of such a wave function in SU(6) $\times$ O(3), 
having a total angular momentum $J$ and projection $J_3$ is given by
\begin{equation}\label{WF}
|\ell S;JJ_3;(\lambda \mu) Y I I_3\rangle  =%}\nonumber \\
\sum_{m_\ell,S_3}
      \left(\begin{array}{cc|c}
	\ell    &    S   & J   \\
	m_\ell  &    S_3  & J_3 
      \end{array}\right) 
       |\ell m_\ell \rangle        
| [f] (\lambda \mu ) Y I I_3; S S_3 \rangle ,
\end{equation}
where the orbital part of the wave function of the entire system, denoted by 
$|\ell m_\ell \rangle$, has a permutation symmetry $[f]$, for simplicity not specified, the same as   
its flavor-spin part $| [f] (\lambda \mu ) Y I I_3; S S_3 \rangle$. 
The wave function  (\ref{WF}), together with an SU$_c(3)$ color singlet $[1^{N_c}]$ forms a totally
antisymmetric $N_c$ quark state.

In order to calculate the expectation value of the mass operator defined in the 
following section one needs to know the matrix elements  of the SU(6) generators $S^i$, $T^a$ and $G^{ia}$ 
between the states  $| [f] (\lambda \mu ) Y I I_3; S S_3 \rangle$ of a given SU(3) symmetry  $(\lambda \mu)$ 
and a spin  $S$, associated to the entire system of $N_c$ quarks. 
In Ref. \cite{Matagne:2008kb} these matrix elements were presented under the form of a 
generalized Wigner-Eckart theorem, containing isoscalar factors of SU(3) and SU(6).
Tables for the most needed isoscalar factors of SU(6) were produced in the same paper 
for the $^28$, $^48$, $^210$ and $^21$ SU(3) $\times$ SU(2) multiplets.
Extended tables were obtained in Ref. \cite{Matagne:2011fr}.

As already mentioned, applications to the  $[{\bf 70},1^-]$ multiplet of the $N = 1$ band
and to the  $[{\bf 70},\ell^-]$ ($\ell$ = 1,2,3) of 
the $N = 3$ band were made in Refs. \cite{Matagne:2011fr} and
\cite{Matagne:2012tm} respectively. In the following we shall follow a similar approach.

%%%%%%%%%%%%%%%%%%%%%%%%%%%%%%%%%%%%%%%%%%%%%%%%%%%%%%%%%%%%%%%%%%%%%%%%%%%
\section{The Mass Operator}

The general form of the mass operator,  where the SU(3) symmetry is broken, has first been proposed in Ref. \cite{JL95} as
\begin{equation}
\label{massoperator}
M = \sum_{i}c_i O_i + \sum_{i}d_i B_i .
\end{equation} 
This is inspired by the perturbative expansion in powers of $1/N_c$ proposed by 't Hooft \cite{HOOFT}
where the operators $O_i$ represent $1/N_c$ corrections to the leading spin-flavor (SF) singlet operator $O_1$
proportional to $N_c$. The contributions of $O_i$ with $i > 1$ estimate the amount of SF symmetry breaking. 
Accordingly, the operators $O_i$ are defined as the scalar products
\begin{equation}\label{OLFS}
O_i = \frac{1}{N^{n-1}_c} O^{(k)}_{\ell} \cdot O^{(k)}_{SF},
\end{equation}
where  $O^{(k)}_{\ell}$ is a $k$-rank tensor in SO(3) and  $O^{(k)}_{SF}$
a $k$-rank tensor in SU(2)-spin, but invariant in SU($N_f$).
Thus $O_i$ are rotational invariant.
For the ground state one has $k = 0$. The excited
states also require  $k = 1$  and $k = 2$ terms.
The rank $k = 1$ tensor has as components the generators $L^i$ of SO(3). 
The components of the $k = 2$ tensor operator of SO(3) are
\begin{equation}\label{TENSOR} 
L^{(2)ij} = \frac{1}{2}\left\{L^i,L^j\right\}-\frac{1}{3}
\delta_{i,-j}\vec{L}\cdot\vec{L},
\end{equation}
which, like $L^i$, act on the orbital wave function $|\ell m_{\ell} \rangle$  
of the whole system of $N_c$ quarks (see  Ref. \cite{Matagne:2005gd} for the normalization 
of $L^{(2)ij}$). 
According to the large $N_c$ counting rules \cite{WITTEN} an $n$-body operator carries a coefficient 
$1/N_c$ reflecting the minimum of $n - 1$ gluon exchanges between two quarks in QCD.
%In practical applications  in the mass formula (\ref{massoperator}) on includes terms up to order $1/N_c$.

The operators $B_i$ break SU(3) explicitly and are defined to have zero expectation
values for nonstrange baryons. Only first order SU(3) breaking terms have been considered so far.

\begin{table}[htb]
\caption{List of dominant operators and their coefficients in the mass formula (\ref{massoperator}) obtained 
in three distinct numerical fits.}
\label{operators}
\renewcommand{\arraystretch}{1.2} % enlarge line spacing
\begin{tabular}{lrrrrr}
\hline
\hline
Operator & Fit 1 &\hspace{0.5cm} & Fit 2 & \hspace{0.5cm} & Fit 3\\
\hline
\hline
$O_1 = N_c \ \1 $               &   616 $\pm$  11     & &  616  $\pm$  11  & &  616  $\pm$  11   \\
$O_2 = \ell^i s^i$            &   150 $\pm$ 239     & &   52  $\pm$  44  & &  243  $\pm$  237            \\
$O_3 = \frac{1}{N_c}S^iS^i$     &   149 $\pm$ 30      & &  152  $\pm$  29 &  &  136  $\pm$   29    \\
$O_4 = \frac{1}{N_c}\left[T^aT^a-\frac{1}{12}N_c(N_c+6)\right]$  &  66 $\pm$ 55 & & 57 $\pm$ 51 & & 86 $\pm$ 55   \\
$O_5 =  \frac{3}{N_c} L^{i} T^{a} G^{i}$                         & -22 $\pm$ 5  & &             & &  -25 $\pm$ 52    \\ 
$O_6 =  \frac{15}{N_c} L^{(2)ij} G^{ia} G^{ja}$                  &  14 $\pm$ 5  & & 14 $\pm$ 5  & &    \\ 
\hline
$B_1 = - \mathcal{S}$           &  23 $\pm$ 38        & &  24 $\pm$ 38    & &  -22 $\pm$ 35     \\     
\hline                  
$\chi_{\mathrm{dof}}^2$         &  0.61               & &   0.52          & &   2.27   \\
\hline \hline
\end{tabular}
\end{table}

%%%%%%%%%%%%%%%%%%%%%%%%%%%%%%%%%%%%%%%%%%%%%%%%%%%%%%%%%%%%%%%%%%%%%%%%%%%%%%%%%%%%%%%%%%%%%%

\begin{table}[htb]
\caption{Matrix elements of $O_i$ for octet resonances.}
\label{NUCLEON}
\renewcommand{\arraystretch}{1.2}
\begin{tabular}{lcccccc}
\hline 
\hline
   &  \hspace{ .3 cm} $O_1$ \hspace{ .3 cm}  & \hspace{ .3 cm} $O_2$  \hspace{ .3 cm} & \hspace{ .3 cm} $O_3$  \hspace{ .3 cm}&  \hspace{ .3 cm} $O_4$  \hspace{ .3 cm} & \hspace{ .3 cm} $O_5$  & \hspace{ .3 cm} $O_6$ \\
\hline
$^48[{\bf 70},2^+]\frac{7}{2}^+$  &  $N_c$   &  $\frac{2}{3}$    & $\frac{15}{4N_c}$ & $\frac{3}{4N_c}$ & $\frac{3(N_c+3)}{2N_c}$ 
&  $-\frac{15(N_c-1)}{4N_c}$\\
$^28[{\bf 70},2^+]\frac{5}{2}^+$  &  $N_c$   &  $\frac{2}{9N_c}(2N_c-3)$ & $\frac{3}{4N_c}$  &   $\frac{3}{4N_c}$ & $\frac{3}{N_c}$ & 0 \\
$^48[{\bf 70},2^+]\frac{5}{2}^+$  &  $N_c$   &  $-\frac{1}{9}$  & $\frac{15}{4N_c}$ &   $\frac{3}{4N_c}$ & $-\frac{N_c+3}{4N_c}$ 
& $\frac{75(N_c-1)}{8N_c}$ \\
$^48[{\bf 70},0^+]\frac{3}{2}^+$  &  $N_c$   &  0    & $\frac{15}{4N_c}$ &   $\frac{3}{4N_c}$  & 0 & 0  \\
$^28[{\bf 70},2^+]\frac{3}{2}^+$  &  $N_c$   &   $-\frac{1}{3N_c}(2N_c-3)$   & $\frac{3}{4N_c}$  &  $\frac{3}{4N_c}$
& $-\frac{9}{2N_c}$ & 0 \\
$^48[{\bf 70},2^+]\frac{3}{2}^+$  &  $N_c$   &   $-\frac{2}{3}$   & $\frac{15}{4N_c}$ &  $\frac{3}{4N_c}$ & $-\frac{3(N_c+3)}{2N_c}$ & 0 \\
$^28[{\bf 70},0^+]\frac{1}{2}^+$  &  $N_c$   &   0   &  $\frac{3}{4N_c}$  & $\frac{3}{4N_c}$   & 0 & 0 \\
$^48[{\bf 70},2^+]\frac{1}{2}^+$  &  $N_c$   &    $-1$   & $\frac{15}{4N_c}$ &  $\frac{3}{4N_c}$ & $-\frac{9(N_c+3)}{4N_c}$
&  $-\frac{105(N_c-1)}{8N_c}$ \\
\hline
\hline
\end{tabular}
\end{table}

%%%%%%%%%%%%%%%%%%%%%%%%%%%%%%%%%%%%%%%%%%%%%%%%%%%%%%%%%%%%%%%%%%%%%%%%%%%%%%%%%%%%%%%%%%%%%%%%%%%%%%

\begin{table}[htb]
\caption{Matrix elements of $O_i$ for decuplet resonances.}
\label{DELTA}
\renewcommand{\arraystretch}{1.2}
\begin{tabular}{lcccccc}
\hline 
\hline
   &  \hspace{ .3 cm} $O_1$ \hspace{ .3 cm}  & \hspace{ .3 cm} $O_2$  \hspace{ .3 cm} &  \hspace{ .3 cm} $O_3$  \hspace{ .3 cm} &  \hspace{ .3 cm} $O_4$  \hspace{ .3 cm} & \hspace{ .3 cm} $O_5$  & \hspace{ .3 cm} $O_6$ \\
\hline

$^210[{\bf 70},2^+]\frac{5}{2}^+$  &   $N_c$   &  $-\frac{2}{9}$ & $\frac{3}{4N_c}$  & $\frac{15}{4N_c}$ & $\frac{3(N_c+1)}{2N_c}$ & 0\\
$^210[{\bf 70},2^+]\frac{3}{2}^+$  &   $N_c$   &  $\frac{1}{3}$  & $\frac{3}{4N_c}$  & $\frac{15}{4N_c}$ & $-\frac{9(N_c+1)}{4N_c}$ & 0 \\
$^210[{\bf 70},0^+]\frac{1}{2}^+$  &   $N_c$   &  0 & $\frac{3}{4N_c}$ & $\frac{15}{4N_c}$ & 0 & 0 \\

\hline
\hline
\end{tabular}
\end{table}

%%%%%%%%%%%%%%%%%%%%%%%%%%%%%%%%%%%%%%%%%%%%%%%%
\begin{table}[htb]
\caption{Matrix elements of $O_i$ for singlet resonances.}
\label{SINGLET}
\renewcommand{\arraystretch}{1.2}
\begin{tabular}{lcccccc}
\hline 
\hline
   &  \hspace{ .3 cm} $O_1$ \hspace{ .3 cm}  & \hspace{ .3 cm} $O_2$  \hspace{ .3 cm} &  \hspace{ .3 cm} $O_3$  \hspace{ .3 cm} &  \hspace{ .3 cm} $O_4$  \hspace{ .3 cm} & \hspace{ .3 cm} $O_5$  & \hspace{ .3 cm} $O_6$ \\
\hline

$^21[{\bf 70},2^+]\frac{5}{2}^+$  &   $N_c$   &  $\frac{2}{3}$ & $\frac{3}{4N_c}$  & $-\frac{2 N_c + 3}{4N_c}$ & $-\frac{N_c-3}{2N_c}$ & 0\\
$^21[{\bf 70},2^+]\frac{3}{2}^+$  &   $N_c$   &  -$1$  & $\frac{3}{4N_c}$  & $-\frac{2 N_c + 3}{4N_c}$ & $\frac{3(N_c-3)}{4N_c}$ & 0 \\
$^21[{\bf 70},0^+]\frac{1}{2}^+$  &   $N_c$   &  0 & $\frac{3}{4N_c}$ & $-\frac{2 N_c + 3}{4N_c}$ & 0 & 0 \\

\hline
\hline
\end{tabular}
\end{table}

%%%%%%%%%%%%%%%%%%%%%%%%%%%%%%%%%%%%%%%%%%%%%%%%%%%%%%%%%%%%%%%%%%%%%%%%%%%%%%%%%%%%%%%%%%%%%%%%%%%%%

Using the experimental data described below we have performed several numerical fits
to obtain the unknown coefficients $c_i$ and $d_i$, which encode the QCD dynamics. As the 
data are still scarce we had to restrict the number of terms in the mass formula,
therefore we had to choose the most relevant operators. They were suggested by our 
previous experience with negative parity states and are exhibited in Table \ref{operators}.

The first is the trivial spin-flavor singlet operator $O_1$ 
of order $\mathcal{O}(N_c)$. 
The first nontrivial operator is the spin-orbit operator $O_2$,
which we identify with the the single-particle operator 
\begin{equation}\label{spinorbit}
\ell \cdot s = \sum^{N_c}_{i=1} \ell(i) \cdot s(i),
\end{equation}
the matrix elements of which are of order $N^0_c$ and are given in Ref. \cite{Matagne:2006zf}
The analytic expression of the matrix elements of $O_2$ can be found in the Appendix A of Ref. \cite{Matagne:2012tm}.

The spin operator $O_3$ and the flavor operator $O_4$ are two-body and linearly independent. 
The expectation value of $O_3$ is  $\frac{1}{N_c} S ( S + 1 )$ where $S$ is the spin 
of the entire system of $N_c$ quarks. 
The expression of the operator $O_4$ given in Table \ref{operators}
is consistent with
the usual $1/N_c (T^a T^a)$ definition in SU(4). In extending it to SU(6) we had to subtract 
the quantity  $(N_c+6)/12$ as explained in Ref. \cite{Matagne:2008kb}. Then, as one can see from  Tables
\ref{NUCLEON}, \ref{DELTA} and \ref{SINGLET},
 the expectation values  of  $O_4$   are positive for octets and decuplets 
and of order  $N^{-1}_c$, as in SU(4), and negative  and of order $N^0_c$  for flavor singlets.

By construction, the operators  $O_5$  and $O_6$ have non-vanishing contributions for orbitally excited states only.
They are also two-body, which means that they carry a 
factor $1/N_c$  in the definition. 
The operator  $O_6$ contains the irreducible spherical tensor (\ref{TENSOR}) and the SU(6)
generator $G^{ja}$ both acting on the whole system.
The latter is a coherent operator which introduces an extra power $N_c$
so that the order of the matrix elements of $O_6$ is $\mathcal{O}(1)$, as it can be seen from Table  
\ref{NUCLEON}. For decuplets and singlets its matrix elements vanish, see Tables  \ref{DELTA} and \ref{SINGLET} respectively.

The matrix elements of $O_5$  and $O_6$ were obtained from the formulas (B2) and (B4) of Ref. \cite{Matagne:2011fr}
where the multiplet $[70,1^-]$ has been discussed.   
The contribution of $O_5$ cancels out for flavor singlets when  $N_c$ = 3,   like 
for $\ell$ = 1 \cite{Matagne:2011fr} and  $\ell$ = 3 \cite{Matagne:2012tm}. 
This property follows from the analytic expression  of the isoscalar factors 
given in Ref. \cite{Matagne:2011fr}. 

Therefore in the mass formula there is one operator, namely $O_1$,  of order $\mathcal{O}(N_c)$
and two operators, $O_2$ and $O_6$ of order $\mathcal{O}(N_c^0)$. They have been used in 
 Refs. \cite{Matagne:2011sn,Matagne:2012vq} where the compatibility of the present approach with 
the meson-nucleon scattering picture has been proven, as mentioned in the introduction.

We remind that the advantage of the present procedure over the standard one, 
where the system is 
separated into a ground state core + an excited quark \cite{CCGL},
is that the number of relevant operators needed in the fit is usually smaller than the number of data and 
it allows a better understanding of their role in the mass formula, in  particular the role of the isospin operator $O_4$ 
which has been omitted in the symmetric core + excited quark procedure in the analysis of mixed symmetric negative
parity states \cite{CCGL,SGS}. 
We should also mention that in our  approach the permutation symmetry remains exact in all applications. % \cite{Matagne:2006dj}.

A comment is in order for the flavor breaking operators $B_i$. 
In the procedure where the system is separated into a core and an excited quark one 
deals with two operators $B_1 = t^8 - \frac{1}{2 \sqrt{3}} $ and 
$B_2 = T^8_c - \frac{N_c - 1}{2 \sqrt{3}} $ 
acting on the excited quark 
and the core respectively (lower case indicates operators acting on the excited quark and subscript $c$ indicates those 
acting on the core). 
These two operators have distinct matrix elements in each sector $^28_J$, $^48_J$, $^210_J$ and $^21_J$ \cite{SGS,Matagne:2006zf}.
In the present method there is a single operator
$T^8 =  t^8  +  T^8_c$ which generates the flavor breaking operator
\begin{equation} 
B_1 = -\frac{2}{\sqrt{3}}(T^8 -\frac{1}{2 \sqrt{3}} \mathcal{O}_1) 
\end{equation}
the matrix element of which is $\langle B_1 \rangle $ = $-\mathcal{S}$, the same for all sectors,
as indicated in Table \ref{operators}, where $\mathcal{S}$ is the strangeness.
Such a result is consistent with Table V of Ref. \cite{Matagne:2006zf} from which we get
\begin{equation}
\langle T^8 \rangle = \frac{N_c + 3 \mathcal{S}}{2 \sqrt{3}},
\end{equation}
for all sectors, as in Ref. \cite{JL95}. This considerably simplifies the situation and implies that the flavor symmetry
breaking picture is different in the present approach as compared to the
symmetric core + excited quark approach, inasmuch as in the first the breaking is independent of the sector and in the
second it is not. This may provide an explanation of the unexpectedly large $\Lambda \Sigma$ splitting obtained in the sector
$^48$,   with the symmetric core + excited quark approach,  see Ref. \cite{Matagne:2006zf},  while presently,
where $\langle B_1 \rangle $ = $-\mathcal{S}$, there is no splitting at all.

%%%%%%%%%%%%%%%%%%%%%%%%%%%%%%%%%%%%%%%%%%%%%%%%%%%%%%%%%%%%%%%%%%%%

{\squeezetable
\begin{table}
%\begin{sidewaystable}
\caption{The partial contribution and the total mass (MeV) predicted by the $1/N_c$ expansion
using Fit 2 of Table \ref{operators}.
The last two columns give  the empirically known masses and the 2012 status in the Review of Particles Properties 
\cite{PDG} .}\label{MASSES}
\renewcommand{\arraystretch}{1.5}
\begin{tabular}{crrrrrrcccl}\hline \hline
                    &      \multicolumn{6}{c}{Part. contrib. (MeV)}  & \hspace{.0cm} Total (MeV)   & \hspace{.0cm}  Exp. (MeV)\hspace{0.0cm}& &\hspace{0.cm}  Name, status \hspace{.0cm} \\

\cline{2-8}
                    &   \hspace{.0cm}   $c_1O_1$  & \hspace{.0cm}  $c_2O_2$ & \hspace{.0cm}$c_3O_3$ &\hspace{.0cm}  $c_4O_4$ &\hspace{.0cm}  $c_6O_6$ & $d_1B_1$&           \\
\hline
$^4N[{\bf 70},2^+]\frac{7}{2}$        & 1848 & 35  & 190 & 14 & -36 & 0    &         $2051\pm44 $  & $2016\pm104$ & & $N(1990)7/2^+$**  \\
$^4\Lambda[{\bf 70},2^+]\frac{7}{2}$  &      &     &    &     &     & 24   &      $2075\pm63 $  & $2094\pm78$  & & $\Lambda(2020)7/2^+$* \\
$^4\Sigma[{\bf 70},2^+]\frac{7}{2}$   &      &     &    &     &     & 24   &      $2075\pm63$  &              & & \\
$^4\Xi[{\bf 70},2^+]\frac{7}{2}$      &      &     &    &     &     & 48   &    $2099\pm93$   &              & & \vspace{0.2cm}\\
\hline
$^2N[{\bf 70},2^+]\frac{5}{2}$        & 1848 & 12  & 38 & 14  & 0   & 0   &   $1912\pm 31$   &$1860\pm70$       & & $N(1860)5/2^+$** \\
$^2\Lambda[{\bf 70},2^+]\frac{5}{2}$  &      &     &    &     &     & 24  &   $1936\pm54$   &              & & \\
$^2\Sigma[{\bf 70},2^+]\frac{5}{2}$   &      &     &    &     &     & 24  &   $1936\pm54$   &              & & \\
$^2\Xi[{\bf 70},2^+]\frac{5}{2}$      &      &     &    &     &     & 48  &   $1959\pm88$  &              & & \vspace{0.2cm}\\
\hline
$^4N[{\bf 70},2^+]\frac{5}{2}$        & 1848 & -6  & 190 & 14 & 89 & 0  & $2136\pm39$   & $2090\pm120$ & & $N(2000)5/2^+$** \\
$^4\Lambda[{\bf 70},2^+]\frac{5}{2}$  &      &     &    &     &   & 24  & $2159\pm60$   & $2112\pm40$  & & $\Lambda(2110)5/2^+$*** \\
$^4\Sigma[{\bf 70},2^+]\frac{5}{2}$   &      &     &    &     &   & 24  & $2159\pm60$  &              & & \\
$^4\Xi[{\bf 70},2^+]\frac{5}{2}$      &      &     &    &     &   & 48  & $2183\pm92$   &              & & \vspace{0.2cm}\\
\hline
$^4N[{\bf 70},0^+]\frac{3}{2}$        & 1848 & 0   & 190 & 14  & 0 &  0  &  $2052\pm18$   & $2052\pm20$  & & $N(2040)3/2^+$* \\
$^4\Lambda[{\bf 70},0^+]\frac{3}{2}$  &      &     &    &     &   & 24  &  $2076\pm49$   &              & & \\
$^4\Sigma[{\bf 70},0^+]\frac{3}{2}$   &      &     &    &     &   & 24  &  $2076\pm49$  &              & & \\
$^4\Xi[{\bf 70},0^+]\frac{3}{2}$      &      &     &    &     &   & 48  &  $2100\pm86$   &              & & \vspace{0.2cm}\\
\hline
$^2N[{\bf 70},2^+]\frac{3}{2}$        & 1848 & -17 &38 & 14  & 0 &  0  & $1883\pm26$   &  $1905\pm30$   & & $N(1900)3/2^+$***\\
$^2\Lambda[{\bf 70},2^+]\frac{3}{2}$  &      &     &    &     &   & 24  & $1907\pm52$   &              & & \\
$^2\Sigma[{\bf 70},2^+]\frac{3}{2}$   &      &     &    &     &   & 24  &   $1907\pm52$   &              & & \\
$^2\Xi[{\bf 70},2^+]\frac{3}{2}$      &      &     &    &     &   & 48  &  $1931\pm87$  &              & & \vspace{0.2cm}\\
\hline
$^4N[{\bf 70},2^+]\frac{3}{2}$        & 1848 & -35 & 190 & 14  & 0 &  0  & $2018\pm30$   &              & & \\
$^4\Lambda[{\bf 70},2^+]\frac{3}{2}$  &      &     &    &     &   & 24  & $2041\pm55$   &              & & \\
$^4\Sigma[{\bf 70},2^+]\frac{3}{2}$   &      &     &    &     &   & 24  & $2041\pm55$  &              & & \\
$^4\Xi[{\bf 70},2^+]\frac{3}{2}$      &      &     &    &     &   & 48  & $2065\pm90$   &              & &
\vspace{0.2cm} \\
\hline
\hline
\end{tabular}
%\end{sidewaystable}{COLEB1}
\end{table}}
%%%%%%%%%%%%%%%%%%%%%%%%%%%%%%%%%%%%%%%%%%%%%%%%%%%%%%%%%%%%%%%%%%%%%%%%%%%%%%%%%%%%%%
{\squeezetable
\begin{table}
\label{multiplet}
\renewcommand{\arraystretch}{1.5}
\begin{tabular}{crrrrrrcccl}\hline \hline
                    &      \multicolumn{6}{c}{Part. contrib. (MeV)}  & \hspace{.0cm} Total (MeV)   & \hspace{.0cm}  Exp. (MeV)\hspace{0.0cm}& &\hspace{-0.2cm}  Name, status \hspace{.0cm} \\

\cline{2-8}
                    &   \hspace{.0cm}   $c_1O_1$  & \hspace{.0cm}  $c_2O_2$ &
		    \hspace{.0cm}$c_3O_3$ &\hspace{.0cm}  $c_4O_4$  &\hspace{.0cm}  $c_6O_6$& $d_1B_1$&     &        \\
\hline
$^2N[{\bf 70},0^+]\frac{1}{2}$        & 1848 & 0   & 38 & 14  & 0  &  0 &   $1900\pm27$   &   & &  \\
$^2\Lambda[{\bf 70},0^+]\frac{1}{2}$  &      &     &    &     &       & 24 &   $1924\pm52$   &            & & \\
$^2\Sigma[{\bf 70},0^+]\frac{1}{2}$   &      &     &    &     &       & 24 &   $1924\pm52$   &   & &  \\ 
$^2\Xi[{\bf 70},0^+]\frac{1}{2}$      &      &     &    &     &       & 48 &   $1948\pm87$  &              & & \vspace{0.2cm}\\
\hline
$^4N[{\bf 70},2^+]\frac{1}{2}$        & 1848 & -52  &190 & 14 &  -125 & 0  &    $1875\pm34$   &  $1870\pm35$ & & $N(1880)1/2^+$** \\
$^4\Lambda[{\bf 70},2^+]\frac{1}{2}$  &      &      &    &    &       & 24 &    $1899\pm58$   &              & & \\
$^4\Sigma[{\bf 70},2^+]\frac{1}{2}$   &      &     &    &     &       & 24 &     $1899\pm58$   &              & & \\
$^4\Xi[{\bf 70},2^+]\frac{1}{2}$      &      &     &    &     &       & 48 &  $1923\pm92$   &              & & \vspace{0.2cm}\\
\hline
$^2\Delta[{\bf 70},2^+]\frac{5}{2}$   & 1848 & -12  & 38  & 72 & 0  &  0 &   $1946\pm58$   & $1892\pm143$ & & $\Delta(2000)5/2^+$**\\
$^2\Sigma'[{\bf 70},2^+]\frac{5}{2}$  &      &     &    &     &    &   24 &    $1970\pm67$   &              & & \\
$^2\Xi'[{\bf 70},2^+]\frac{5}{2}$     &      &     &    &     &    &   48 &    $1994\pm92$  &              & & \\
$^2\Omega[{\bf 70},2^+]\frac{5}{2}$   &      &     &    &     &    &   71 &    $2018\pm124$  &              & &\vspace{0.2cm} \\
\hline
$^2\Delta[{\bf 70},2^+]\frac{3}{2}$   & 1848 & 17  & 38  & 72 & 0  & 0   &   $1975\pm64$ &                &  & \\
$^2\Sigma'[{\bf 70},2^+]\frac{3}{2}$  &      &     &    &     &    & 24   &  $1999\pm71$ &                &  & \\
$^2\Xi'[{\bf 70},2^+]\frac{3}{2}$     &      &     &    &     &    & 48   &   $2023\pm95$&                &  & \\
$^2\Omega[{\bf 70},2^+]\frac{3}{2}$   &      &     &    &     &    & 71   &   $2046\pm126$&                &  & \vspace{0.2cm}\\
\hline
$^2\Delta[{\bf 70},0^+]\frac{1}{2}$   & 1848 &  0  & 38 &  72 &  0 &  0  &   $1958\pm59$ &   &  &  \\
$^2\Sigma'[{\bf 70},0^+]\frac{1}{2}$  &      &     &    &     &    & 24   &  $1982\pm68$ &  $1896\pm95$   &  & $\Sigma(1880)1/2^+$** \\
$^2\Xi'[{\bf 70},0^+]\frac{1}{2}$     &      &     &    &     &    & 48   &   $2005\pm93$                 &  & \\
$^2\Omega[{\bf 70},0^+]\frac{1}{2}$   &      &     &    &     &    & 71   &    $2029\pm124$&                &  & \vspace{0.2cm}\\
\hline
$^2\Lambda'[{\bf 70},2^+]\frac{5}{2}$ & 1848 &  35 & 38   & -43   & 0   & 24  &   $1901\pm84$ &                &  &\vspace{0.2cm} \\
\hline
$^2\Lambda'[{\bf 70},2^+]\frac{3}{2}$ & 1848 & -52 & 38   & -43    & 0 &  24  &   $1815\pm87$ &                &  & \vspace{0.2cm}\\
\hline
$^2\Lambda'[{\bf 70},0^+]\frac{1}{2}$ & 1848 &  0  & 38   & -43    & 0 &  24  &  $1867\pm77$ &  $1791\pm64$   &  & $\Lambda(1810)1/2^+$*** 
\vspace{0.2cm}\\
\hline
\hline
\end{tabular}
%\end{sidewaystable}
\end{table}}

%%%%%%%%%%%%%%%%%%%%%%%%%%%%%%%%%%%%%%%%%%%%%%%%%%%%%%%%%%%%%%%%

\section{The experimental situation} 

In our previous work \cite{Matagne:2006zf} we have made use of the Baryon Particle Listings of the Particle Data Group  
before 2012 and made averages over the baryon masses of the Karlsruhe-Helsinki group \cite{hoehler79}  and the
Carnegie Mellon-Berkeley group  \cite{Cutkosky:1980rh}
or considered some values obtained by Manley and Saleski  \cite{Manley:1992yb}. 
Here we rely on the 2012 version of the Review of Particle Properties (PDG) \cite{PDG} which incorporates the new 
multichannel partial wave analysis of the Bonn-Gatchina group  \cite{Anisovich:2011fc}. 
The changes in PDG for positive parity resonances (for a summary of the 
Bonn-Gatchina group see Ref. \cite{Klempt:2012gu}) are important for our work.

First, the resonance $P_{13}(1900)$ has been upgraded from two to three stars with a Breit-Wigner mass of
 1905 $\pm$  30 MeV.
Second, the resonance  $N(2000)5/2^+$ 
has been split into two two-star resonances $N(1860)5/2^+$ and $N(2000)5/2^+$fo
with masses  indicated in Table \ref{MASSES}. The suggestion was that $N(1860)5/2^+$ belongs to a quartet
\cite{Anisovich:2011su}.
There is a new one-star resonance $N(2040)3/2^+$ observed in the decay $J/\psi \rightarrow p \bar p \pi^0$.
There is also a new two-star resonance $N(1880)1/2^+$  observed by the Bonn-Gatchina group with a mass of
1870 $\pm$ 35 MeV \cite{Anisovich:2011fc},
which confirms a previous observation by  Manley and Saleski  \cite{Manley:1992yb} where a mass 
of 1885 $\pm$ 30 MeV has been found.
%%%%%%%%%%%%%%%%%%%%%%%%%%%%%%%%%%%%%%%%%%%%%%%%%%%%%%%%%%%%%%%%%%% 

\section{Results and discussion}
We have performed several numerical fits for finding the unknown coefficients $c_i$ and $d_i$ of the mass formula (\ref{massoperator}) 
using the 2012 Review of Particle 
Properties (PDG) \cite{PDG} which incorporates the new 
multichannel partial wave analysis of the Bonn-Gatchina group  \cite{Anisovich:2011fc}, implying the  
changes described in Sec. IV.  In Table \ref{operators} we present three of the most favorable fits.

Actually we have started by including all experimentally known resonances located in the appropriate mass region,
except  for those which were supposed to belong to the $[\bf{56},2^+]$ multiplet \cite{GSS03}. Finally we found out that 
only a selective choice of resonances give a reasonable fit when described by the formalism presented above.  

The final result includes 11 resonances, having a status of three, two or occasionally one star.
There are no 4-star resonances as candidates for the $[\bf{70},\ell^+]$ multiplet. The selection 
we have made is described below. 
As experimental masses  we took either the Bonn-Gatchina group results,
or we averaged over all values indicated in the Particle Listings of PDG \cite{PDG}.
For example, for $\Delta(2000)5/2^{+**}$ and $\Sigma(1880)1/2^{+**}$ we averaged over three and eleven experimental 
values respectively.

As a matter of fact we have included the new $N(1860)5/2^{+**}$ and $N(2040)3/2^{+*}$ resonances 
%Although assigned to a $[\bf{70},2^+]$ multiplet,  
and obtained a better numerical fit when interpreting the $N(1860)5/2^{+**}$ resonance as a member of a spin doublet 
(see Table \ref{MASSES}) instead of a quartet, 
as proposed in Ref. \cite{Anisovich:2011su}. The reason is that the spin operator $O_3$ contributes 
with a quantity proportional to $S(S + 1)$
 and $c_3$ is positive, see Table \ref{operators}, 
so that a doublet member should be below a spin quartet member with $J^P = 5/2^+$. The latter is thus expected  
to have a mass larger than 1860 MeV.  
However we agree with Ref. \cite{Anisovich:2011su} that the resonance $N(1880)1/2^{+**}$ belongs to a spin quartet,
see Table \ref{MASSES}.

On the other hand, in order to obtain  natural sizes for the coefficient $c_i$ \cite{Pirjol:2003ye},
from the final fit we have removed several resonances which were included
in our previous work based on the excited quark + symmetric core procedure \cite{Matagne:2006zf},
but which are not compatible with the present approach.  
These are the $N(1710){1/2^{+***}}$ and the 
$\Sigma(1770){1/2^{+*}}$ resonances. 
The theoretical argument is that their masses are too low. 
On the experimental side one can justify the removal of  the $N(1710)1/2^{+***}$ resonance as due to
the latest GWU analysis  of Arndt et al. \cite{Arndt:2006bf} where it has not been seen. 
This is anyhow a controversial resonance.

We had also ignored
the $\Delta(1750)1/2^{+*}$ resonance, considered previously \cite{Matagne:2006zf}, inasmuch as,
neither Arndt et al. \cite{Arndt:2006bf} nor Anisovich et al.  \cite{Anisovich:2011fc} find evidence
for it.

From Table \ref{operators} one can see that  $\chi_{\mathrm{dof}}^2$ 
of Fits 1, 2 and 3 are  0.61,   0.52   and    2.27 respectively.
In  Fit 1 the mass formula contains the operators up to 
order $1/N_c$ included, which, according to our previous experience  
with mixed symmetric negative parity states, see, for example, \cite{Matagne:2012tm},
are thought to be the most dominant. 
Note that despite a good $\chi_{\mathrm{dof}}^2$, the coefficient $c_2$ of the spin-orbit
operator is not well determined. Its central value is consistent with predictions from 
the $1/N_c$ expansion for the $N$ = 1 band \cite{Pirjol:2003ye} but we expect
a smaller $c_2$ in the $N$ = 2 band, inasmuch as the contribution from the spin-orbit 
operator decreases with the excitation energy  \cite{Matagne:2005gd}.

In Fit 2 we have removed the operator $O_5$ and obtained a reasonable value for $c_2$.
In Fit 3 we have removed the operator $O_6$, which like  $O_2$ is
of order $N^0_c$, crucial for the compatibility of the quark-shell 
picture used here and the more fundamental meson-nucleon scattering picture,
as discussed in Ref.  \cite{Matagne:2012vq}.
In Fit 3 the coefficient $d_1$ of the SU(3) breaking term becomes negative which is not good
for the mass sequence within a multiplet.
Thus the presence of $O_6$ is necessary. It implies once more that the model used here is compatible 
with the contracted SU(2$N_f$) symmetry which is exact
when $N_c \rightarrow \infty$. 

Thus Fit 2 is the best fit.
The baryon masses calculated from this fit with the formula (\ref{massoperator})
are exhibited in Table \ref{MASSES}, together with the partial contributions of various operators. 
One can clearly see that the isospin operator $O_4$, neglected in the 
symmetric core + excited quark studies of the $N$ = 1 band is crucial for the fit.
To the masses of the decuplet members it  contributes nearly two times more than the spin operator $O_3$. 
As its matrix elements are negative for flavor singlets, see Table \ref{SINGLET}, it also allows a good description
of the  $\Lambda(1810)1/2^+$  resonance.  The important role of $O_4$ is
in agreement with the conclusion of previous studies on negative parity states 
\cite{Matagne:2011fr,Matagne:2012tm}.

%%%%%%%%%%%%%%%%%%%%%%%%%%%%%%%%%%%%%%%%%%%%%%%%%%%%%%%%
% Section VI

\section{Regge trajectories}
In Ref. \cite{Matagne:2005gd} we searched for a systematic global behavior 
of some $c_i$ coefficients as a function of the excitation energy,
$\emph i. e.$ as a function of the band number $N$. 
Accordingly, we have plotted some of the known  $c_i$ at that time
for $N \leq 4$. The points corresponding to mixed symmetric states    
were obtained from the symmetric core + excited quark approach. 
There were no studies of the $N$ = 3 band available yet. We found that
$c_1$ increases linearly as a function of $N$, while $c_2$  and the spin term
coefficient decrease as a function of $N$, as expected from quark models.

These findings inspired further studies to establish a connection between the $1/N_c$ expansion 
method and a simple semi-relativistic quark model with a Y-junction confinement potential plus
a hyperfine interaction generated by one gluon exchange,  
both for nonstrange and strange baryons \cite{Semay:2007cv,Semay:2007ff}.
The band number $N$ emerged naturally from both approaches. 
We found that the large $N_c$ results for $c^2_1$ are practically 
indistinguishable from the quark model results and they followed a linear Regge trajectory as a function of $N$. 
The linear Regge trajectories are a manifestation of the non-perturbative aspect of QCD dynamics, which at
long distance becomes dominated by confinement \cite{Karch:2006pv}.
Indeed, let us denote by $M_{qqq}$ the contribution of the kinetic plus the confinement energy in
the quark model. Then from the identification of this contribution with the leading spin-flavor
singlet operator of the large $N_c$ mass formula one has
\begin{equation}
c^2_1 = M^2_{qqq}/9,
\end{equation}
where we have set $N_c$ = 3,  so that the values of $c^2_1$ were compared to the quark model results,
see Fig. 1 of Ref. \cite{Buisseret:2008tq} where a review can also be found.

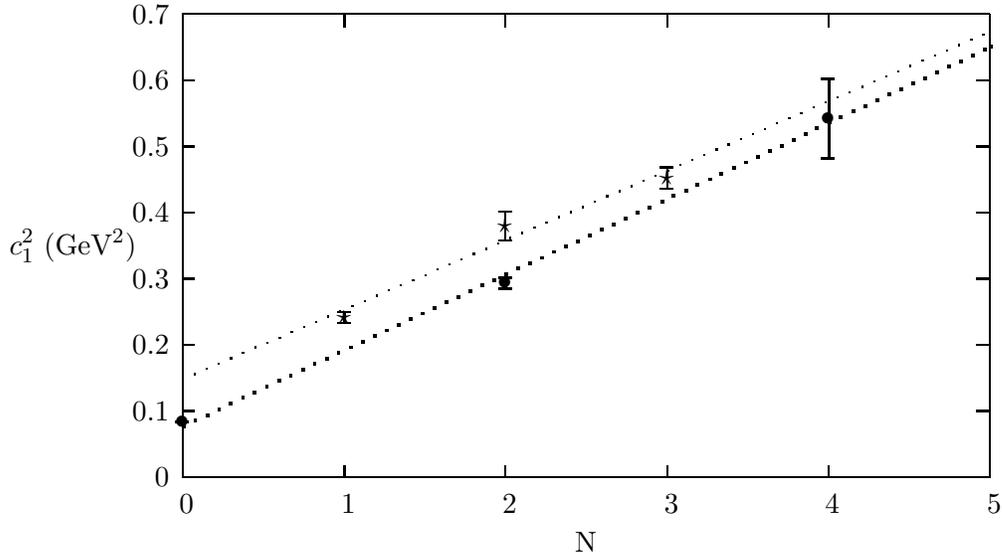
\begin{figure}
\begin{center}
% GNUPLOT: LaTeX picture
\setlength{\unitlength}{0.240900pt}
\ifx\plotpoint\undefined\newsavebox{\plotpoint}\fi
\sbox{\plotpoint}{\rule[-0.200pt]{0.400pt}{0.400pt}}%
\begin{picture}(1500,900)(0,0)
\sbox{\plotpoint}{\rule[-0.200pt]{0.400pt}{0.400pt}}%
\put(171.0,131.0){\rule[-0.200pt]{4.818pt}{0.400pt}}
\put(151,131){\makebox(0,0)[r]{ 0}}
\put(1419.0,131.0){\rule[-0.200pt]{4.818pt}{0.400pt}}
\put(171.0,235.0){\rule[-0.200pt]{4.818pt}{0.400pt}}
\put(151,235){\makebox(0,0)[r]{ 0.1}}
\put(1419.0,235.0){\rule[-0.200pt]{4.818pt}{0.400pt}}
\put(171.0,339.0){\rule[-0.200pt]{4.818pt}{0.400pt}}
\put(151,339){\makebox(0,0)[r]{ 0.2}}
\put(1419.0,339.0){\rule[-0.200pt]{4.818pt}{0.400pt}}
\put(171.0,443.0){\rule[-0.200pt]{4.818pt}{0.400pt}}
\put(151,443){\makebox(0,0)[r]{ 0.3}}
\put(1419.0,443.0){\rule[-0.200pt]{4.818pt}{0.400pt}}
\put(171.0,547.0){\rule[-0.200pt]{4.818pt}{0.400pt}}
\put(151,547){\makebox(0,0)[r]{ 0.4}}
\put(1419.0,547.0){\rule[-0.200pt]{4.818pt}{0.400pt}}
\put(171.0,651.0){\rule[-0.200pt]{4.818pt}{0.400pt}}
\put(151,651){\makebox(0,0)[r]{ 0.5}}
\put(1419.0,651.0){\rule[-0.200pt]{4.818pt}{0.400pt}}
\put(171.0,755.0){\rule[-0.200pt]{4.818pt}{0.400pt}}
\put(151,755){\makebox(0,0)[r]{ 0.6}}
\put(1419.0,755.0){\rule[-0.200pt]{4.818pt}{0.400pt}}
\put(171.0,859.0){\rule[-0.200pt]{4.818pt}{0.400pt}}
\put(151,859){\makebox(0,0)[r]{ 0.7}}
\put(1419.0,859.0){\rule[-0.200pt]{4.818pt}{0.400pt}}
\put(171.0,131.0){\rule[-0.200pt]{0.400pt}{4.818pt}}
\put(171,90){\makebox(0,0){ 0}}
\put(171.0,839.0){\rule[-0.200pt]{0.400pt}{4.818pt}}
\put(425.0,131.0){\rule[-0.200pt]{0.400pt}{4.818pt}}
\put(425,90){\makebox(0,0){ 1}}
\put(425.0,839.0){\rule[-0.200pt]{0.400pt}{4.818pt}}
\put(678.0,131.0){\rule[-0.200pt]{0.400pt}{4.818pt}}
\put(678,90){\makebox(0,0){ 2}}
\put(678.0,839.0){\rule[-0.200pt]{0.400pt}{4.818pt}}
\put(932.0,131.0){\rule[-0.200pt]{0.400pt}{4.818pt}}
\put(932,90){\makebox(0,0){ 3}}
\put(932.0,839.0){\rule[-0.200pt]{0.400pt}{4.818pt}}
\put(1185.0,131.0){\rule[-0.200pt]{0.400pt}{4.818pt}}
\put(1185,90){\makebox(0,0){ 4}}
\put(1185.0,839.0){\rule[-0.200pt]{0.400pt}{4.818pt}}
\put(1439.0,131.0){\rule[-0.200pt]{0.400pt}{4.818pt}}
\put(1439,90){\makebox(0,0){ 5}}
\put(1439.0,839.0){\rule[-0.200pt]{0.400pt}{4.818pt}}
\put(171.0,131.0){\rule[-0.200pt]{0.400pt}{175.375pt}}
\put(171.0,131.0){\rule[-0.200pt]{305.461pt}{0.400pt}}
\put(1439.0,131.0){\rule[-0.200pt]{0.400pt}{175.375pt}}
\put(171.0,859.0){\rule[-0.200pt]{305.461pt}{0.400pt}}
\put(0,495){\makebox(0,0){$c^2_1$ (GeV$^2$)}}
\put(805,29){\makebox(0,0){N}}
\put(425.0,374.0){\rule[-0.200pt]{0.400pt}{4.095pt}}
\put(415.0,374.0){\rule[-0.200pt]{4.818pt}{0.400pt}}
\put(415.0,391.0){\rule[-0.200pt]{4.818pt}{0.400pt}}
\put(678.0,503.0){\rule[-0.200pt]{0.400pt}{11.081pt}}
\put(668.0,503.0){\rule[-0.200pt]{4.818pt}{0.400pt}}
\put(668.0,549.0){\rule[-0.200pt]{4.818pt}{0.400pt}}
\put(932.0,584.0){\rule[-0.200pt]{0.400pt}{8.191pt}}
\put(922.0,584.0){\rule[-0.200pt]{4.818pt}{0.400pt}}
\put(425,383){\makebox(0,0){$\star$}}
\put(678,526){\makebox(0,0){$\star$}}
\put(932,601){\makebox(0,0){$\star$}}
\put(922.0,618.0){\rule[-0.200pt]{4.818pt}{0.400pt}}
\put(171,285){\usebox{\plotpoint}}
\put(171.00,285.00){\usebox{\plotpoint}}
\put(190.20,292.86){\usebox{\plotpoint}}
\put(209.24,301.11){\usebox{\plotpoint}}
\put(228.26,309.41){\usebox{\plotpoint}}
\put(247.28,317.67){\usebox{\plotpoint}}
\put(266.21,326.17){\usebox{\plotpoint}}
\put(285.17,334.62){\usebox{\plotpoint}}
\put(304.37,342.48){\usebox{\plotpoint}}
\put(323.53,350.43){\usebox{\plotpoint}}
\put(342.49,358.87){\usebox{\plotpoint}}
\put(361.46,367.29){\usebox{\plotpoint}}
\put(380.65,375.15){\usebox{\plotpoint}}
\put(399.79,383.15){\usebox{\plotpoint}}
\put(418.64,391.79){\usebox{\plotpoint}}
\put(437.72,399.95){\usebox{\plotpoint}}
\put(456.91,407.81){\usebox{\plotpoint}}
\put(475.92,416.13){\usebox{\plotpoint}}
\put(494.91,424.50){\usebox{\plotpoint}}
\put(514.00,432.62){\usebox{\plotpoint}}
\put(533.15,440.58){\usebox{\plotpoint}}
\put(552.14,448.90){\usebox{\plotpoint}}
\put(571.15,457.21){\usebox{\plotpoint}}
\put(590.26,465.27){\usebox{\plotpoint}}
\put(609.30,473.52){\usebox{\plotpoint}}
\put(628.40,481.62){\usebox{\plotpoint}}
\put(647.41,489.93){\usebox{\plotpoint}}
\put(666.54,497.94){\usebox{\plotpoint}}
\put(685.27,506.87){\usebox{\plotpoint}}
\put(704.41,514.88){\usebox{\plotpoint}}
\put(723.60,522.74){\usebox{\plotpoint}}
\put(742.57,531.16){\usebox{\plotpoint}}
\put(761.53,539.59){\usebox{\plotpoint}}
\put(780.69,547.55){\usebox{\plotpoint}}
\put(799.88,555.44){\usebox{\plotpoint}}
\put(818.76,563.99){\usebox{\plotpoint}}
\put(837.77,572.30){\usebox{\plotpoint}}
\put(856.95,580.21){\usebox{\plotpoint}}
\put(875.99,588.46){\usebox{\plotpoint}}
\put(895.03,596.70){\usebox{\plotpoint}}
\put(914.04,605.01){\usebox{\plotpoint}}
\put(933.14,613.07){\usebox{\plotpoint}}
\put(952.25,621.12){\usebox{\plotpoint}}
\put(971.27,629.41){\usebox{\plotpoint}}
\put(990.30,637.67){\usebox{\plotpoint}}
\put(1009.41,645.71){\usebox{\plotpoint}}
\put(1028.50,653.81){\usebox{\plotpoint}}
\put(1047.51,662.12){\usebox{\plotpoint}}
\put(1066.56,670.33){\usebox{\plotpoint}}
\put(1085.60,678.58){\usebox{\plotpoint}}
\put(1104.44,687.28){\usebox{\plotpoint}}
\put(1123.64,695.14){\usebox{\plotpoint}}
\put(1142.66,703.44){\usebox{\plotpoint}}
\put(1161.65,711.79){\usebox{\plotpoint}}
\put(1180.73,719.95){\usebox{\plotpoint}}
\put(1199.87,727.93){\usebox{\plotpoint}}
\put(1218.88,736.19){\usebox{\plotpoint}}
\put(1237.89,744.50){\usebox{\plotpoint}}
\put(1256.99,752.61){\usebox{\plotpoint}}
\put(1276.03,760.86){\usebox{\plotpoint}}
\put(1295.15,768.90){\usebox{\plotpoint}}
\put(1314.16,777.21){\usebox{\plotpoint}}
\put(1333.27,785.28){\usebox{\plotpoint}}
\put(1352.31,793.53){\usebox{\plotpoint}}
\put(1371.41,801.62){\usebox{\plotpoint}}
\put(1390.42,809.93){\usebox{\plotpoint}}
\put(1409.43,818.21){\usebox{\plotpoint}}
\put(1428.57,826.19){\usebox{\plotpoint}}
\put(1439,831){\usebox{\plotpoint}}
\sbox{\plotpoint}{\rule[-0.400pt]{0.800pt}{0.800pt}}%
\put(171,218){\usebox{\plotpoint}}
\put(161.0,218.0){\rule[-0.400pt]{4.818pt}{0.800pt}}
\put(161.0,218.0){\rule[-0.400pt]{4.818pt}{0.800pt}}
\put(678.0,427.0){\rule[-0.400pt]{0.800pt}{4.095pt}}
\put(668.0,427.0){\rule[-0.400pt]{4.818pt}{0.800pt}}
\put(668.0,444.0){\rule[-0.400pt]{4.818pt}{0.800pt}}
\put(1185.0,632.0){\rule[-0.400pt]{0.800pt}{30.112pt}}
\put(1175.0,632.0){\rule[-0.400pt]{4.818pt}{0.800pt}}
\put(171,218){\makebox(0,0){$\bullet$}}
\put(678,436){\makebox(0,0){$\bullet$}}
\put(1185,695){\makebox(0,0){$\bullet$}}
\put(1175.0,757.0){\rule[-0.400pt]{4.818pt}{0.800pt}}
\sbox{\plotpoint}{\rule[-0.500pt]{1.000pt}{1.000pt}}%
\put(171,211){\usebox{\plotpoint}}
\put(171.00,211.00){\usebox{\plotpoint}}
\put(189.85,219.70){\usebox{\plotpoint}}
\put(208.52,228.76){\usebox{\plotpoint}}
\put(227.35,237.47){\usebox{\plotpoint}}
\put(245.86,246.85){\usebox{\plotpoint}}
\put(264.58,255.79){\usebox{\plotpoint}}
\put(283.30,264.75){\usebox{\plotpoint}}
\put(302.15,273.45){\usebox{\plotpoint}}
\put(320.99,282.15){\usebox{\plotpoint}}
\put(339.81,290.90){\usebox{\plotpoint}}
\put(358.50,299.92){\usebox{\plotpoint}}
\put(377.35,308.62){\usebox{\plotpoint}}
\put(396.19,317.32){\usebox{\plotpoint}}
\put(414.85,326.39){\usebox{\plotpoint}}
\put(433.70,335.09){\usebox{\plotpoint}}
\put(452.55,343.79){\usebox{\plotpoint}}
\put(471.31,352.66){\usebox{\plotpoint}}
\put(490.05,361.56){\usebox{\plotpoint}}
\put(508.90,370.26){\usebox{\plotpoint}}
\put(527.74,378.96){\usebox{\plotpoint}}
\put(546.41,388.03){\usebox{\plotpoint}}
\put(565.25,396.73){\usebox{\plotpoint}}
\put(584.10,405.43){\usebox{\plotpoint}}
\put(602.81,414.40){\usebox{\plotpoint}}
\put(621.61,423.20){\usebox{\plotpoint}}
\put(640.45,431.90){\usebox{\plotpoint}}
\put(659.30,440.60){\usebox{\plotpoint}}
\put(678.04,449.52){\usebox{\plotpoint}}
\put(696.81,458.37){\usebox{\plotpoint}}
\put(715.65,467.07){\usebox{\plotpoint}}
\put(734.50,475.77){\usebox{\plotpoint}}
\put(753.16,484.84){\usebox{\plotpoint}}
\put(772.00,493.54){\usebox{\plotpoint}}
\put(790.85,502.24){\usebox{\plotpoint}}
\put(809.54,511.27){\usebox{\plotpoint}}
\put(828.36,520.01){\usebox{\plotpoint}}
\put(847.20,528.71){\usebox{\plotpoint}}
\put(866.00,537.50){\usebox{\plotpoint}}
\put(884.71,546.48){\usebox{\plotpoint}}
\put(903.56,555.18){\usebox{\plotpoint}}
\put(922.40,563.88){\usebox{\plotpoint}}
\put(941.07,572.95){\usebox{\plotpoint}}
\put(959.91,581.65){\usebox{\plotpoint}}
\put(978.76,590.35){\usebox{\plotpoint}}
\put(997.60,599.05){\usebox{\plotpoint}}
\put(1016.27,608.12){\usebox{\plotpoint}}
\put(1035.11,616.82){\usebox{\plotpoint}}
\put(1053.96,625.52){\usebox{\plotpoint}}
\put(1072.73,634.36){\usebox{\plotpoint}}
\put(1091.46,643.29){\usebox{\plotpoint}}
\put(1110.31,651.99){\usebox{\plotpoint}}
\put(1129.15,660.69){\usebox{\plotpoint}}
\put(1147.39,670.56){\usebox{\plotpoint}}
\put(1166.23,679.26){\usebox{\plotpoint}}
\put(1185.08,687.96){\usebox{\plotpoint}}
\put(1203.80,696.90){\usebox{\plotpoint}}
\put(1222.58,705.73){\usebox{\plotpoint}}
\put(1241.43,714.43){\usebox{\plotpoint}}
\put(1260.27,723.14){\usebox{\plotpoint}}
\put(1278.94,732.20){\usebox{\plotpoint}}
\put(1297.78,740.90){\usebox{\plotpoint}}
\put(1316.63,749.60){\usebox{\plotpoint}}
\put(1335.47,758.30){\usebox{\plotpoint}}
\put(1354.14,767.37){\usebox{\plotpoint}}
\put(1372.98,776.07){\usebox{\plotpoint}}
\put(1391.83,784.77){\usebox{\plotpoint}}
\put(1410.53,793.76){\usebox{\plotpoint}}
\put(1429.34,802.54){\usebox{\plotpoint}}
\put(1439,807){\usebox{\plotpoint}}
\sbox{\plotpoint}{\rule[-0.200pt]{0.400pt}{0.400pt}}%
\put(171.0,131.0){\rule[-0.200pt]{0.400pt}{175.375pt}}
\put(171.0,131.0){\rule[-0.200pt]{305.461pt}{0.400pt}}
\put(1439.0,131.0){\rule[-0.200pt]{0.400pt}{175.375pt}}
\put(171.0,859.0){\rule[-0.200pt]{305.461pt}{0.400pt}}
\end{picture}
\end{center}
\caption{The coefficient $c^2_1$ (GeV$^2$) as a function of the band number $N$. The numerical values of $c_1$ were 
taken from Ref. \cite{Semay:2007cv} for $N$ = 0, from Ref. \cite{Matagne:2011fr} Fit 3 for $N$ = 1,
from Ref. \cite{GSS03} for $N$ = 2 $[{\bf 56},2^+]$, from the present work Fit 2 for $N$ = 2 $[{\bf 70},\ell^+]$ ($\ell$ = 0,2),
from Ref. \cite{Matagne:2012tm} Fit 3 for $N$ = 3 $[{\bf 70},\ell^-]$ ($\ell$ = 1,2,3), from Ref. \cite{MS1} for $N$ = 4 $[{\bf 56},4^+]$.
The heavy dots refer to $[56]$-plets and the stars to $[70]$-plets.
The best fit of these data was obtained with two distinct linear trajectories. }
\end{figure}

Presently, we have a consistent description of mixed symmetric positive and 
negative parity states corresponding  to $N$ = 1, 2 and 3 bands. 
It is interesting to revisit the
the Regge trajectory problem. In Fig. 1 we plot  $c^2_1$ as a function of the band number $N$
for $N \leq 4$. The value of $c_1$ at $N$ = 3 is presently known \cite{Matagne:2012tm},
while in Ref. \cite{Matagne:2005gd} the corresponding point was missing.
One can see that two distinct trajectories emerge from this new picture,
one for symmetric ${\bf[56]}$-plets, the other for mixed symmetric ${\bf[70]}$-plets. 
This behavior reminds that obtained in Ref. \cite{Goity:2007sc} where two distinct trajectories 
have been found for the evolution of $(N_c c_1)^2$ as a function of the angular momentum $\ell \leq 6$
(Chew-Frautschi plots).  Note that in Ref. \cite{Goity:2007sc} the mixed symmetric states were
described within the the ground state core + excited quark 
approach. The mass operator was reduced  to the contribution of
the $\mathcal{O}(N_c)$ spin-flavor singlet, 
the $\mathcal{O}(1/N_c)$ hyperfine spin-spin interaction, acting between core quarks only,
and SU(3) breaking terms.  As a consequence there is 
no contribution from the spin dependent terms in flavor singlets because their core has $S_c$ = 0.
There are no $\mathcal{O}(N^0_c)$ contributions.
For a consistent treatment, in Ref.  \cite{Goity:2007sc}  the hyperfine interaction was restricted to
core quarks in symmetric states as well. It was not necessary to specify whether or not the core is excited,
due to the simplicity of the mass operator.
  
In our case, the symmetric and mixed symmetric states are treated on an equal basis:
there is no distinction between the core and an excited quark (the core may be excited as well), 
the Pauli principle is always fulfilled and all quark-quark interaction terms are included. 
The existence of two distinct Regge trajectories, 
one for symmetric, another for mixed symmetric states, 
may be due to the existence
of terms of order $N^0_c$ in the mass formula for mixed symmetric states,
which often bring a negative contribution, see ${\emph e. g.}$ the operator $O_6$,
while for symmetric states the expansion starts at order $1/N_c$.
This may require the coefficient $c_1$ to be larger for 
mixed  symmetric states, for compensating the negative contribution 
of  operators of order $N^0_c$.

%%%%%%%%%%%%%%%%%%%%%%%%%%%%%%%%%%%%%%%%%%%%%%%%%%%%%%%%
% Section VII

\section{Conclusions}
We have revisited the mass spectrum of resonances supposed to belong to the $[{\bf 70, \ell^+}]$ multiplets of the $N$ = 2 band
with $\ell$ = 0 or 2 in the light of a recent multichannel partial wave analysis which enriched
the Review of Particle Properties in 2012. We found that the new resonances can well be described as
belonging to the above multiplets. However, we found more appropriate to describe the resonance $N(1860)5/2^+$ 
as a member of a doublet rather than that of a spin quartet, at variance with the suggestion of Ref. \cite{Anisovich:2011su}.  
The three-star resonance $N(1710){1/2^{+***}}$  does not fit into our treatment of
the $[{\bf 70, \ell^+}]$ multiplets. It would be useful to better 
understand its nature. 

We point out that the $1/N_c$ expansion method  allows us to search for a classification 
of excited baryons into SU(6) x O(3) multiplets, as presently shown in Table  \ref{MASSES}. 
This is a natural and useful extension of the classification of ground state baryons. 
It allows us to make predictions for the mass range of unknown baryons as members of octets 
or decuplets, which may guide experimentalists in the search for highly excited or strange 
baryons for which data are scarce. 

Like for the $N$ = 1 and 3 bands, we found that both the quark spin and isospin operators,
acting on the entire system,  play dominant roles
in describing the data. 
In the symmetric core + excited quark approach applied to the $N$ = 1 band \cite{CCGL} the
latter was split into the core part $T^2_c$ (where its contribution is identical to that of
the spin part $S^2_c$) and a term, $\frac{1}{N_c} t^a T^a_c$ acting between the excited quark
and the core, which was ignored.  In our previous work \cite{Matagne:2006zf} we found that $\frac{1}{N_c} t^a T^a_c$
contributes with some important amount to the mass. 

Two distinct Regge trajectories have been found for symmetric and mixed symmetric states. 
An important remark is that the bases of operators used for different mixed symmetric multiplets was the same,
irrespective of the resonance parity.
The extension of our studies to resonances belonging to $N > 4$ would help in better understanding 
the existence of two Regge trajectory, although we are aware that the number of experimental data 
is very limited at higher energies.

%%%%%%%%%%%%%%%%%%%%%%%%%%%%%%%%%%%%%%%%%%%%%%%%%%%%%%%%%%%%%%%%%%%%%%%%%%%%

\end{document}